# Hash Chain Links Resynchronization Methods in Video Streaming Security: Performance Comparison

Emad Abd-Elrahman, Mohamed Boutabia and Hossam Afifi
TELECOM SudParis, CNRS SAMOVAR UMR 5157
9, Rue Charles Fourier - 91011 Evry Cedex, France
{emad.abd_elrahman, mohamed.boutabia, hossam.afifi} @it-sudparis.eu

*Abstract*—Hash chains provide a secure and light way of security to data authentication including two aspects: Data Integrity and Data Origin Authentication. The real challenge of using the hash chains is how it could recover the synchronization state and continue keeping the hash link in case of packet loss? Based on the packet loss tolerance and some accepted delay of video delivery which are representing the permitted tolerance for heavy loaded applications, we propose different mechanisms for such synchronization recovery. Each mechanism is suitable to use according to the video use case and the low capabilities of end devices. This paper proposes comparative results between them based on the status of each one and its overhead. Then, we propose a hybrid technique based Redundancy Code (RC). This hybrid algorithm is simulated and compared analytically against the other techniques (SHHC, TSP, MLHC and TSS). Moreover, a global performance evaluation in terms of delay and overhead is conducted for all techniques.

## I. INTRODUCTION

THE transmission of multimedia applications over Internet occupies a wide band of research. The challenging points like security and reliability are representing the whole part of interest in applications like video streaming and IPTV delivery based security measures. Many security mechanisms were proposed for securing the delivery of real time applications based on Hash Chains methodology. Hash chains are very popular security mechanism for securing many applications such as authentication of multicast traffic [1, 8], routing of sensor networks and sensors applications [2], privacy of RFID authentication [3], data streaming [4], micropayment systems [5], one time password [6] and many data origin authentication applications. The main advantage of hash chains is the light calculations compared to other cryptographic algorithms like the encryption methods. It also provides a fast and secure way for the real time applications that are very sensitive to any delay caused by the security overhead.

### A. Video Streaming Security Measures

According to the recommendations by National Institute of Standards and Technology (NIST) [16] for securing sensitive applications and also for defining the degree of security, there are four levels of security. These levels were organized based on group of roles define the co-relation between the operators and the provided services. That standard provides four increasing, qualitative levels of security: *Level 1* (basic security measures), *Level 2* (high physical layer security), *Level 3* (Identity-Based security and more services authorization), and *Level 4* (high level security applications).
For video streaming, the objective is different because the security module or in general the cryptographic module must take into account the nature of this application (time sensitive application), and the quality of delivery affected by the security measures delay. So, we can build our prospection to secure video streaming on the degree of importance of this stream and also the capabilities of the low end devices that will be used by the clients to access this stream like PDA devices.

Fig. 1 illustrates the security measures classification for our video streaming study. We divide the security modules into two levels according to the nature of video diffusion (online or offline video stream) and who diffused it (important speech or normal speech). We propose a suitable hashing technique for each level based on hash chains mechanism as shown in Fig. 1.

Our solution adapts to different cases like when the stream has a big important but is not online. In this scenario, we will apply a cross layer security mechanism between the two levels shown in the figure.

Therefore, the applied security measure for video streaming will not be fixed for all types of streams but it will vary according to the video requirements, the video status and the network conditions.

We may combine hashing and watermarking so as to assure a high degree of security. These measures could be used and combined with our hash chain methodology according to the type of applications under security as follows:

1. **Digital Rights Management (DRM):** is mainly designed to prevent illegal accessing, copying or converting of multimedia materials into other formats using digital devices. DRM is a generic term for access control technologies that used copyright protections. Signatures and watermarks are classes of DRM.
2. **Cryptographic Signature:** used for authentication purposes like detection of any alteration of the signed data and to authenticate the data sender.

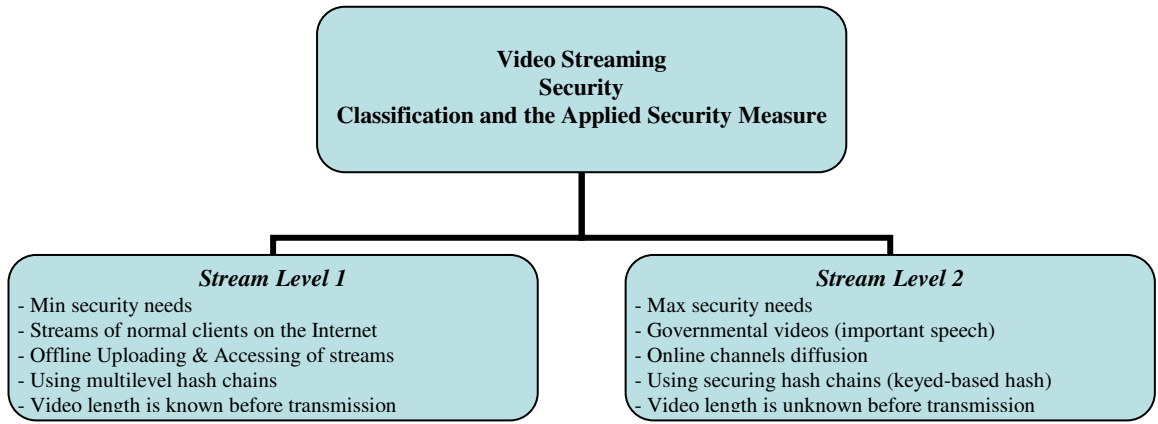

Fig. 1: video streaming security levels classification.

3. **Watermarks:** are used for authentication in especial applications and are designed to resist alterations and modifications in data.
4. **Fragile Watermarks:** are watermarks that have only very limited robustness. They are used to detect modifications of the watermarked data (like image applications).

The ability to achieve good security for real time applications requires some security measures from the above items merged with the hashing mechanism.

Using a hash function is a simple way to ensure data confidentiality. A hash function transforms a string of characters into a usually shorter, fixed-length value or key that represents the original string. The difference between hashing and encryption is in how the data is stored. With encrypted mode, the data can be decrypted with a key. With the hash functions, after the data is entered and converted using the hash function, the plaintext is gone. Therefore, the hashed values are only used in comparison. We have two base standards for hash building as follows:

- *SHA-1* [20]: When a message of any length $< 2^{64}$ bits is input, the SHA-1 produces a 160-bit (20 Bytes) output called a message digest. So, $2^{160}$ operations are needed for knowing the digest value (the number of possibilities that can be generated with 160 bit length).
- *MD5* [19]: When a message of any length $< 2^{64}$ bits is input, the MD5 produces a 128-bit (16 Bytes) output called a message digest. So, $2^{128}$ operations are needed for knowing the digest value (the number of possibilities that can be generated with 128 bit length).

For simplicity, our work relies on using SHA-1 and comparing the overhead with MD5 as they used the same block-based calculations. But, the proposed algorithm could accommodate any type of new hashing like SHA-256 or SHA-512. Although, there are a lot of critiques to MD5 like the trials to break it within max 1 hour proposed in [25], it is still implemented in many security applications. Also, the US declared that, it will gradually change to high series of SHA for the government applications. More collision resistance analysis about SHA family is listed in [26]. Actually, our objective is the reliability by finding solution to the hash links in hash chains technique. So, some details about attack mitigations will be shown in Section IV.

Practically, we have two cases of video streaming as follows:

**Offline Video Streaming:** The video is in this case on the server side and has a definite length. So, the server could calculate any security measure for that total length before starting the client accessing it. YouTube and Dailymotion are good examples for this category of video sharing servers which have a huge database of short videos [13]. The Video-on-Demand (VoD) is representing this case study.

**Online Video Streaming:** This scenario is more complex. The length of video file in this case is unknown, so the sender can not calculate any measures of security when receiving from the up-loader or diffusion at different time periods. This scenario represents the personal TV or live video. It corresponds also to video streaming channels hosted by some Internet content providers.

*B. Related Work*

Hash chain is a successive application of a cryptographic hash function *h(.)* to any string. The link of chain means that; the initialization value currently input to *h(.)* will be the output of the previous hash calculated from the previous part of data. So, the hash calculations could not continue if the previous hash digest missed.

Hash chains for video streaming have been considered extensively in the literature. The survey in [23] provides a good study about using hash chain in video streaming. It mainly conducted a comparison between many algorithms which proposed handling the issues of hash chain with video stream authentications. However, handling of the resynchronization problem for broken hash links still needs additional work. Our previous work [17] highlighted the resynchronization issue in hash chain links and categorized some solutions for it. Then, we added some security measures based on signing specific packets in the video stream [18]. The work in [9] gives a good starting point for how to sign a digital streaming video. Authors proposed two cases; offline and online streams. They chain blocks based on the packets inside the block. Each block carries the hash of the next one (online case). For the offline case they calculate the hash based on the whole video and the receiver must have some buffer so as to start the verification after a specific length of the video. Their algorithm does not handle the redundancy of chain links.

In [10], authors introduce the Butterfly Graph. They divided the packets into groups and each group has one signature calculated based hashing. The redundancy is achieved by sending the signed packets several times. Their overall concern is to keep a good performance as the amount of redundancy is increased. Also, in [31,32] they examine the problem of streaming of authenticated video over lossy public networks depending on the ideas of Graph and taking into account the quality of wireless channels. It is a kind of optimization technique for authenticating the streaming packets which called Rate-distortion-Authentication (R-D-A). Moreover, they achieved remarkable optimization in media quality and packet overhead.

In [11], the work is based on signing a small number of special packets in data stream; each packet is linked to a signed packet via multiple hash chain. The links depend on six hashes per packet. Hence six packets carry the same hash value and this represents a large overhead. Two solutions for securing the video stream are compared. The first solution is called TESLA (Timed Efficient Stream Loss-tolerant Authentication). The second scheme is called EMSS (Efficient Multichained Stream Signature).

Another work [12], handled the video stream authentication by assuming a combination of one-way hashes and digital signatures to authenticate packets. Their idea can be explained as follows; for collision resistant, the hash of packet $P_i$ is appended to packet $P_{i+1}$ before signing $P_{i+1}$, then the signature on $P_{i+1}$ guarantees the authenticity of $P_i$ and $P_{i+1}$ at the same time. The drawback of that proposal is the large overhead as it increased linearly with the growth numbers of packets.

A time-critical multicast authentication scheme was proposed in [27], which combines hash chains with one time signature to authenticate streaming of packets. The algorithm provides short end-to-end computational latency, perfect tolerance to packet loss, and strong resistance against malicious attacks. They used long key for achieving high security which leads to large overhead.

*C. Work Motivations and Organization*

Our objective in this paper is to study and design novel solutions for hash chain resynchronization in case of some packets loss. For conducting this study, we assume some parameters that will be repeated in many sections as shown in Table I in Section III.

This work will completely focus on the handling mechanisms for the re-initialization problem to keep the continuous hash chain in case of packet loss. This loss can break the link of hash chains and may lead to restart the process again. We propose adding some redundancy codes that will be calculated based on the hash values of different *Blocks* from the video. Those redundancy values will be inserted in some packets inside the *Block*. Hence, those values will help the receiver side to extract the hash values for each Block without correctly received the whole packets of this *Block*. This means that, in case of some packets are lost from the *Block*, this loss will not affect on the continuity of hash chain used to authenticate the video transmission and also will not lead to stop the streaming.

The rest of this work is organized as follows: Part II gives overview on the hash chain synchronization problem and the proposed solutions comparison. Part III illustrates our algorithm architecture and assumptions for achieving redundancy of hash link. Some attacks analyses are studied in Section IV. Section V shows the results and Section VI concludes our work and its future directions.

II. HASH CHAINS RESYNCHRONIZATION

Hash chain is an old technique used in many applications. Our work depends on a simple forward one way hashing system as shown in Fig. 2. This type of sequential hashing is often used in practice and requires relatively less memory than other types of parallel hashing. Moreover, it is more convincible for low capacity end devices. If we adopt the traditional hash chain in its typical way it will request a complex synchronization system. So, it will need large memory and buffering capacities from the clients. So, the adopting of forward and sequential way in hashing system will avoid these difficulties. Also, the buffering sizes will depend on our redundancy factors for how many Blocks/Window under processing as we will explain later in results section.

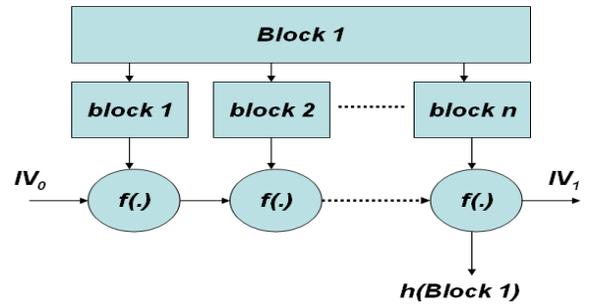

Fig. 2: Simple construction of hash chain mechanism for processing one Block/Window

**$IV_0$:** is the standard Initialization Vector according to the type of hash algorithm
**$IV_1$:** is the final output from hashing Block 1 and will be the initialization vector of Block 2
**block:** is representing standard unit for processing hash function (for example 512 bits for MD5 or SHA-1 Modulo 512)
**Block:** is representing a group of packets (for example one Block=100 packets)

When the hash chains are applied to the video streams like VoD or IPTV, it faces several problems for keeping the hash link continuity in case of packet drops. As some packets are lost, this will cause mismatch calculation in hash link or Message Digest (MD) between sender and receiver. So, we are searching for continuity of hash chain in case of that loss happened.

In the next sections we categorize the solutions for this problem into four categories (SHHS, TSP, MLHC and TSS). Then, we discuss the advantages and the disadvantages of each technique. Moreover, we suggest a new hybrid technique that collects the best features from the four methods of hash link synchronisation.

### A. Self-Healing Hash Chain (SHHS)

Self-Healing is a modern technique that is used a lot in the smart networks. It means that creating a process that has the ability to recover its state in case of system failures without external help. The technique was used a lot with hash chain in many applications like ''Self-Healing Key Distribution'' [15] which focus on the users capability to recover the lost group keys on their own, without requesting additional transmissions from the group manager.

Video streaming can be delivered based on TCP or UDP transport protocols. TCP is mainly used to overcome Network Address Translation (NAT) filters but, the most appropriate is UDP. Usually, during online streaming, we have some packet drops that can be considered with respect to some packet loss tolerance. The acceptable tolerance may not affect the streaming quality. However, the loss could affect on synchronization of the hash chains of the stream.

The Self-Healing Hash Chain (SHHC) can overcome this problem by re-synchronizing the chains despite the loss of some packets from the stream. The SHHC is a robustness system able to resolve the synchronisation problem of chains. As in the Fig. 3; the stream is divided into specific time blocks and at each time a hash must be calculated for this period of time. The reliability will depend on the redundancy factor of the concatenated hashes. To guarantee the synchronisation three hashes may be concatenated together.

This procedure will add some redundancy for tracking the link synchronization points of the stream. Also, the advantage of this scheme is the low overhead for memory and calculations.

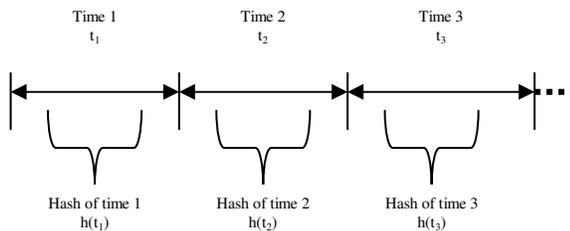

Fig. 3: The time hash-chain of sending party

As in the Fig. 3; the stream is divided into specific time *Blocks* and for each period of time a hash must be calculated. The sending parity will depend on some redundancy of the concatenated hash values that will be transmitted. We strongly recommend that three hashes must be concatenated so as to guarantee the synchronisation of links between sender and receiver. The concatenated values are inserted in the last packet of each *Block* of the video.

The sequence for concatenation mode can be explained as follows:
- Hash of time zero will be *sign* || $h(t_1)$, means the signed packet of the first part.
- Hash of $t_1$ time-end will be $h(t_1)$ || $h(t_2)$
- Hash of $t_2$ time-end will be $h(t_1)$ || $h(t_2)$ || $h(t_3)$
- Hash of $t_3$ time-end will be $h(t_2)$ || $h(t_3)$ || $h(t_4)$
- Hash of $t_n$ time-end of stream will be $h(t_{n-1})$ || $h(t_n)$

The pros of concatenations are: concatenating outputs from multiple hash functions provides collision resistance as good as the strongest of the algorithms included in the concatenated result.

For less overhead, we can replace the concatenation process by XORing process. This replacement will reduce the overhead sent with in the stream by 1/3 for the base of 3-hash concatenated together.

The self-healing feature comes from the receiver ability to extract or recalculate any hash without receiving the total Block or Window of packets.

So, at any time the sender transmit three hashes to link the time Block of this time with the previous time Block and the coming one. This procedure will add some redundancy for tracking the synchronisation points of the stream.

### B. Time-Synchronization Point (TSP)

This mechanism is used to assure synchronisation of hash chains in case of packet loss. It depends on adding additional information bits to the stream. The stream must be divided into a pre-defined specific time Blocks. After each Block, a synchronisation point must be inserted in the sender side as shown in Fig. 4. Those inserted points can help the receiver tracking the synchronisation of the stream.

In this case the receiver must keep in tracking those time synchronisation points (TSP) so as not to lose the stream synchronisation or hash link breaks. The drawback of this technique is the large bits overhead added for synchronisation because it depends on adding an extra packet for this purpose. For more information about overhead comparison see Table II.

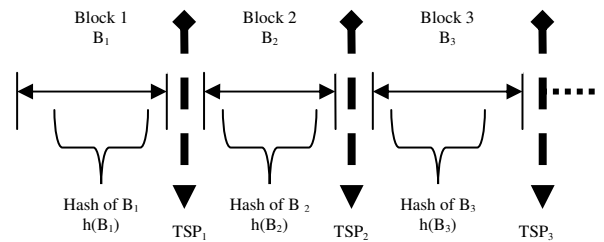

Fig. 4: The time synchronisation point of sending party

In this method, if we assume that each *Block* or *Window* has (*N*) packets, then the packet number (*N+1*) will be redundant packet. As, we insert this packet after each Block, it will add a global overhead on the stream depends on the total number of Blocks in the video file.

## C. Multi Layer Hash Chain (MLHC)

This technique was used for some applications and gave good results for the problem of security assurance for the E-lottery winners and their serial numbers generation tickets [7]. The multi layer means here to have calculations for hash chains where each calculation is representing one layer according to the base of calculation. Although the objectives are different (e-lottery and video streaming), this technique could be very effective especially in offline video streaming security mode. When we use this technique in video streaming the layers conception will completely be different so as to match the specific nature of the real time applications. We can highlight the impacts of two layers hash by the example in Fig. 5. In this structure, we have two concurrent layers of hashing as:

1. $H`_i = h(W_i, IV_{st})$: unkeyed hashing step which depends on standard initialization vector
2. $H_i = h(H`_i, IV_{sec})$: keyed hashing step where the key is equal the $IV_{sec}$ ( secure IV)

And if the round function used is $h(.)$, then:

$H`_i = h(h(......h(h(IV_{st}, W_1), W_2), ......... W_{i-1}), W_i)$
$H_i = h(h(......h(h(IV_{sec}, H`_1), H`_2), ......... H`_{i-1}), H`_i)$

This nested double layer hash chain can thwart many high level attacks to the stream and the hashing itself.

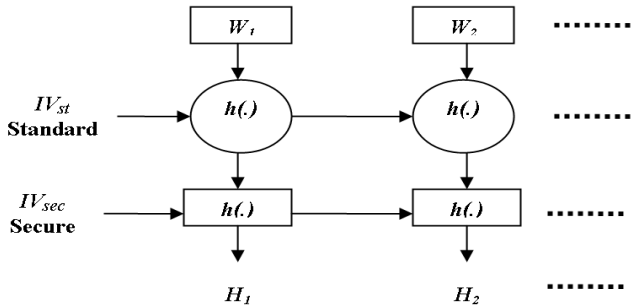

Fig. 5: Two layer hash chain mechanism

## D. TimeStamp Synchronization (TSS)

The sequence of packets could be used as a good measurement for achieving video synchronization and also keeping the link of hash chain. This can be efficient with less calculation cost and time overhead because the timestamp is a mandatory field in Real Time Protocol (RTP) packet as described in [14] for RTP packets of MPEG-4 streams. In this case, The RTP packets are responsible for sequence numbers and timestamp synchronisation (TSS) between the source and destination. The benefit of this technique is the reuse of parameters from RTP standard packets header. Therefore, the sequence number and timestamp for each packet are good indication or index to where is the lost point. So, the added digest value plus the original packet timestamp in all stream packets or in one packet per Block or Window of packets represent the link between sequential hash chain outputs.

| V=2|P|X| CC |M| PT | sequence number |
|---|---|
| Timestamp | |
| Synchronization source (SSRC) identifier | |
| contributing source (CSRC) identifiers | |
| .... | |
| MPEG-4 stream (byte aligned) | |
| ...OPTIONAL RTP padding | |

Fig. 6: RTP packet for MPEG-4 stream as in [14]

*Window* time-stamp: the time stamp that will help in hash chain link synchronisation is 4 bytes per packet as shown in the Fig. 6. But, for our *Window* based of calculations, we will consider the 4 bytes only overhead per *Window* for achieving hash link synchronization which nothing added else the digest value according the cryptographic hash algorithm used. Therefore, if we have two parities under synchronization of timestamp, then the last packet of each Window or what is called the Window timestamp packet will responsible of hash link synchronization as the following:

Let $T_i$ is the timestamp of Window $W_i$, then the *Window Digest* will be:

$$WD_i = h(W_i, H_{i-1})$$

where $H_{i-1}$ is previous Window Digest $W_{i-1}$. But, we need to add the timestamp to this calculation by concatenating it to the previous Digest. So, the final Window hash will be:

$$H_i = h( WD_i \| T_i , H_{i-1})$$

So, the hash chain links relations $H_{i+1}, H_i, H_{i-1}$ could be built based on the *Windows timestamps* $T_{i+1}, T_i , T_{i-1}$.

## E. The Hybrid Technique Based (RC)

Finally, we suggest a hybrid technique capable of inheriting the advantages of the previous solutions and also overcoming most of their drawbacks. This technique could be useful for slow processor endpoints so as to minimise the calculation needs to be performed during every session. Also, it could be used to rapidly re-establish the link synchronization in case of packet loss session problems or delay time. Another criterion is packet and hash information caching that have more CPU intensive compared with using time synchronization that may be important for low-memory mobile platforms.

Our proposal is inspired from redundancy code techniques. **Redundancy Code (RC)** is a generic concept introducing some redundancy in the system to overcome hash chain break in case of packet loss and to increase the reliability of the security system. In the next Section, we introduce more details about using RC with hash chain.

## III. THE PROPOSED REDUNDANT HASH CHAIN METHOD

Before describing the proposed architecture, we must have a look on the packetization sequence of the video stream. The high part of Fig. 7 illustrates the simple sequence in standard manner based on real time transport protocol [14]. The video is considered as a group of chunks output from the coder such as MPEG-TS [24]. This gives better clarification on which the stream word represents for us and what is the packet structure for our proposal. The hashing calculations will be done after the first row of Fig. 7 (for transmitter) and before in the case of the receiver.

The proposed architecture in Fig. 7 has many parameters that need to be initialized:

**Stream of Chunks:** are the output blocks after MPEG-TS (like RTP packets).

**Blocks $B_i$:** is a group of packets that have a relation with their numbered Chunks. For example; each Block=10 Packets and each Packet=7 Chunks in case of RTP Packet.

**$IV_0$:** is the initial victor for starting the hash chain.

**h(.):** is the hash functions used for calculate the output hash like MD5 or SHA series.

**$h_i$:** is the output hash value or the output digest.

**Hi**: is the output hash digest for two layer hash technique.

**Combination Code:** is the coding process that will be used to calculate a redundancy code for generating the hash value in case of missing a hash value of Block (ex. XOR function).

**$RC_i$:** is the output Redundancy Code that is responsible for recalculate the missed hash value so as to keep the hash link not broken. It is something like Forward Error Correction (FEC) codes.

The added Redundancy Code will allow the receiver to detect and correct the errors in the digest values (under some restrictions). This code is the key factor for solving the resynchronization problem of hash link.

Hash link recovery could happen without asking the sender for additional data retransmission because the sender is memory-less in case of *online* video streaming. The advantages of RCs are that; buffering is not required and the retransmission of hash values can often be avoided (which reduces the bandwidth requirements, time calculations and the buffering memory). RC is therefore applied in this situation where the retransmissions are difficult to achieve in real time applications and memory-less devices. The main objective from RCs is the hash link synchronization and finding the recovery point of synchronization by obtaining the hash value of this time which represents the initial vector (IV) for next hash calculation in our chain.

*A. Assumptions*

All the previous proposals mentioned in section III differentiated between the concept of *offline* and *online* video streaming. They made their calculations based on the pre-known video length in case of *offline stream*. Also, the *online stream* has an infinite length assumption. In both cases, if the receiver has some restrictions about the processing capabilities and the buffering capacities (Memory Buffers), the two cases lead to one case only which is the *online* scenario.

Our proposed solution is built to fit the new generation of handheld devices that have some limitations in all processing capabilities compared to the normal PCs. So, the treatment of any video will be considered as an *online* one (from the receiver side) although if in some cases the sender knows all videos lengths accessed by the others. This assumption will eliminate the need of buffering of data at the receiver side before starting the playing of video in case of large videos.

We assumed that, the redundancy in this case is mandatory for synchronization matter. But, when we calculate the RC for some part of data, this calculation will mainly depend on the degree of redundancy and the accepted overhead.

For example, if RC calculated based on 3 hashes values like $RC_1$= combination ($H_1$, $H_2$, $H_3$) and $RC_2$= combination ($H_3$, $H_4$, $H_5$) then we have redundancy 3/4 with dynamic sliding Window. But, if we consider $RC_1$= combination ($H_1$, $H_2$, $H_3$) and $RC_2$= combination ($H_4$, $H_5$, $H_6$) then, it will represent the static sliding Window which means that no relation between the two Windows. If we take 4 hashes values the redundancy will be 4/5 and so on. So, which factor will be control the calculation of the RCs codes? This is one of the most effective factors in the calculations.

In Table I, we assumed some parameters and values that we used in the calculation of hashing and RC values. All of the assumed parameters in the table were preselected based on the packet standardization size for Real Time Protocol (RTP). The calculated sizes for the Block and the Window are output result from the analytical and simulation results based Matlab.

TABLE I
PROPOSING PARAMETERS USED

| Parameters | Symbol | Definition |
|---|---|---|
| Packet | P | *Standard packet size like MTU size of 1500 Bytes* |
| block | b | *Standard block size for hashing algorithm like 512 bits for MD5 or SHA-1* |
| Block | B | *The number of packets to be processed together* |
| Packet Rate | PR | *= VBR/MTU (packets per sec)* |
| Block Rate | BR | *= PR/ Block Size (Blocks per sec)* |
| Video Bit Rate | VBR | *For example 512 Kbps or 1 Mbps* |
| Window | W | *Is dynamic buffer contains number of Blocks* |
| Hash Function | h(.) | *Is the hash algorithm used like MD5 or SHA-1* |
| Hash Output | H | *Is the output digest or hash value of each Block* |
| Packet Error Rate | PER | *Probability of packets loss or error in the Block* |
| Hash Error Rate | HER | *HER = (PER/Block size).RF this for any Window* |
| Redundancy Factor | RF | *The number of Blocks per Window processing in scanning* |

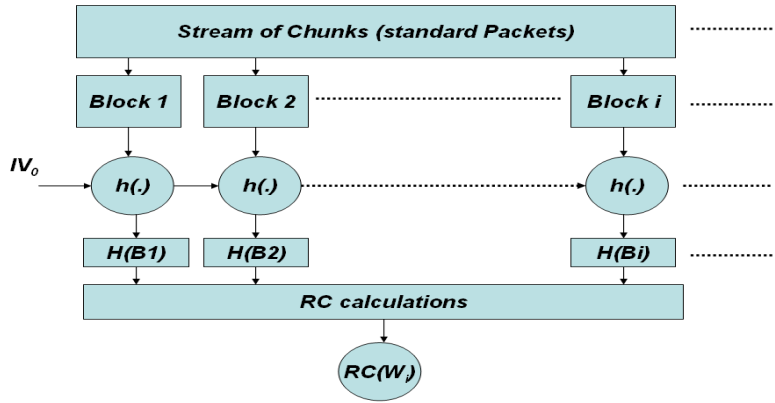

Fig. 7: Block diagram for the hash chain redundancy for video streaming. The original stream is divided into chunks (series of packets) then assembled them to specific Blocks $B_i$ after that the hash chain applied to the Blocks; finally, the RCs calculated based on static Window size.

### B. Sender Security Tasks

This part focuses on how the sender prepares the packets and puts the calculated hash values and redundancy code in the packets? Also, how each packet will have an index to that place in the packet which caries this redundancy code?

The Window mechanism adopts a technique of dynamic buffering. This buffer depends on some parameters like video rate and client processing capabilities. After an agreement done between the sender and receiver, the sliding *Window* mechanism will be conducted according to the redundancy factors adopted.

The complete steps for the implemented algorithm are:
1. Input video file.
2. Divide the file into chunks by MPEG-TS coder each (188 Bytes).
3. Define the Packet size (each packet 7 chunks).
4. Define the Block size (variable from 10 to 100).
5. Define the used hash function MD5 or SHA-1.
6. Start hashing Block by Block with initialization vector of current Block is the hash value of previous Block (chain mechanism).
7. Calculate the Redundancy Code (XOR two or 3 hash together).
8. Insert the RC code in specific packet (or more than one).
9. Index each packet with the location of RC place [17].
10. Add transport headers and send the packets of Block according to the Window size.

### C. Receiver Verification Tasks

All treatments of the received packets are after RTP layer. The receiver will handle the verification of *Blocks* or *Windows* according to the *Window* size which controlled by the Redundancy Factor (RF). Therefore, it can process one Block and compare its hash digest with the received one. If they are identical, this means that the received Block is correct and it will process the next. Otherwise, it will wait till the *Window* complete its size and then use the RC value to drive the hash of the previous *Block* to help the receiver continuing its verifications for the next Block.

The total procedure for verification is as follows:

1. Read the received RTP packets.
2. group the Block size (ex. from 10 to 100 packets)
3. verify the whole Block secured hash and the index for each packet
4. compare the receive hash with the calculated one
5. drop the packets that not have the correct index
6. use the redundancy code RC in case of packet loss to know the hash value or the signature of that Block
7. divide the packets to chunks for MPEG decoder
8. decode the packet elements of the video
9. run the application to view the video in case of ( OK) for the predefined tolerance for packet loss
10. return verification pass (OK)

### D. The Recovery Time

The recovery time is the receiver waiting time before recovering the missed hash link based on the RC value. This time must be less than the standard RTT value.

We have two scenarios for recovery:

**Best Case:** delay time for recovery is very small in comparing to Round Trip Time (RTT) to avoid requesting new IV for reinitialization process.

**Worst Case:** delay time will be larger than the best case because the loss happened in the beginning of the *Window* and the receiver will wait some times till receive the entire *Window*. But, in this case almost the delay will be less than RTT or the receiver will prefer reinitializing than recovering.

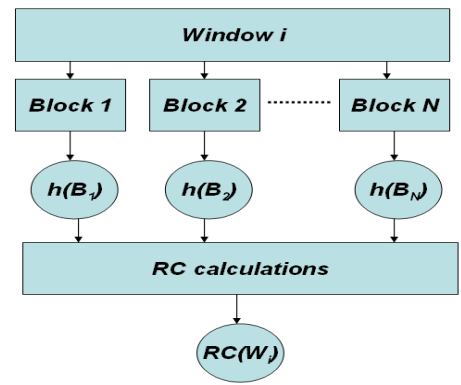

Fig. 8: processing of static *Window* sliding over N *Blocks*

If we define the Recovery Time as the waiting time for the receiver to recover the hash link in case off any Block error. As shown in Fig. 8, any Window consists of (N) Blocks and the Block Rate is BR Block/sec as shown in Table I. So we can calculate the waiting time as:

Waiting Time: $W_T = (N-i)/BR$ where (i) is the error Block position in the Window

For the best case: (i = N), the error occurred in the last Block of the Window (ideal scenario for recovery)

$$W_{Tb} = (N-i)/BR = N-N/BR = Zero$$

For the worst case: (i=1), the error occurred in the first Block

$$W_{Tw} = (N-i)/BR = (N-1)/BR$$

In all cases the $W_{Tb}$ or $W_{Tw}$ must be less than the **RTT** value so as to prove that; it is best for the receiver to depend on RC value for recovering any missed hash link rather than requesting reinitialization IV from the sender.

### E. Offline Access Initialization

As in the Fig. 9 below, we have two phases: the uploader-server phase and the client-server access phase. We assume that; the first phase is pre-secured by the server side. Moreover, this phase can be secured more and more using encryption techniques especially in this offline scenario as the online real time feature not exist.

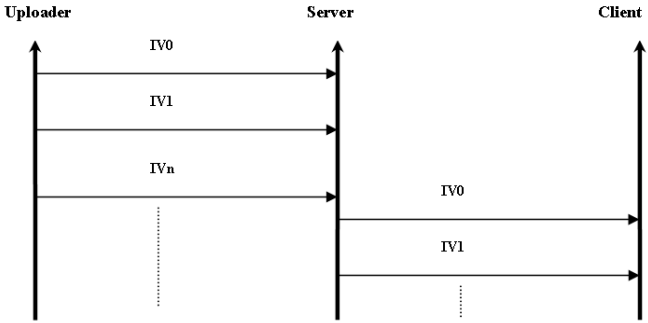

Fig.9: offline joining case

*Client joins procedure*

When a user wants to join the media streaming server, he/she should first pass the authentication phase in the secured manner as it explained in [18]. Then, it will be assigned directly to the first *Window* number and starting the indexing from zero ($IV_0$) because there is no need for its timestamp as the access is *offline* and must start from the beginning of the video.

### F. Online access initialization

At any time, an uploader can start his online video diffusion and any user can access this stream from the hosting server at the time instant of his joining. The server initializes the $IV_0$ for this client by $IV_t$ where t is current time of the server side.

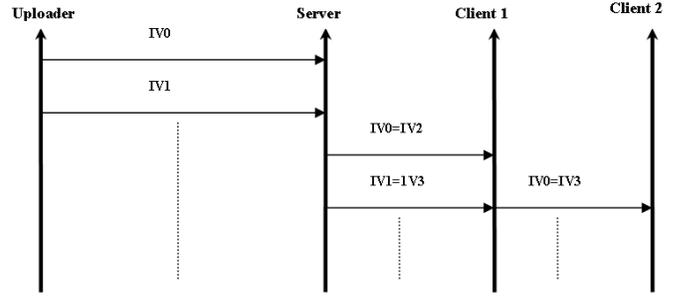

Fig. 10: online case with different joining time access for clients

In Fig. 10, client 1 joined the online stream at time $t_2$ which means that; he missed 2 *Windows* from the beginning of the stream and client 2 joined at $t_3$ which means he missed 3 *Windows* from the starting time of diffusion.

*Client joins procedure*

When a user wants to join the media streaming server, he/she should first pass the authentication phase in the secured manner explained in [18]. Then, it can find the trusted starting point according to its time-stamp for assigning to the nearest *Window* index number [17].

### G. Security Exchange Phase

This phase focuses on the key agreement between the server and clients. Also, its objective is to generate either a secure (IV) to be used in hash chain or a secure private key that will use to sign the hash value. There are many security algorithms that can handle this process like Diffie-Hellman (DH) [28] or Elliptic Curve Cryptography (ECC) [29]. Fig.11 illustrates all the steps needed in this phase. We use an elliptic curve key agreement based on Identity Based Cryptography (IBC) [30]. As each user has a YouTube email address (for example: bob@gmail.com), we will use it to generate a public key. This will lead to a personalized key agreement to access to YouTube services.

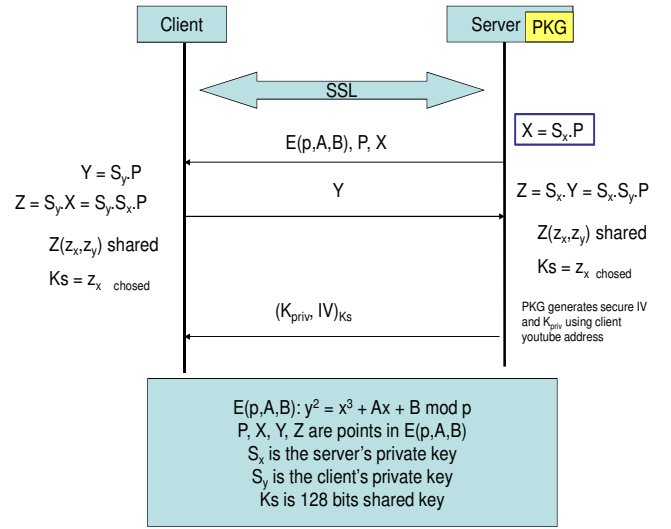

Fig.11: Security exchange phase

In this phase, there are two important steps. The first one is the key agreement to generate a shared key (Ks) and the second one is the generation of client private key $K_{priv}$ by the Private Key Generator (PKG). The two steps need to be realized in a secure manner.

After the login access, an SSL (Secure Sockets Layer) session starts. The server chooses an elliptic curve defined in Galois Field GF(p) where p is a 128 bits prime number. This elliptic curve has A and B like coefficient. It will be defined as E(p,A,B): $y^2 = x^3 + Ax + B \mod p$. The server chooses a public point P in E and computes the public point $X = S_x.P$ where $S_x$ is the server private key. E(p,A,B), P and X are sent to the client as shown in Fig.11.

This latter calculates a point $Y = S_y.P$ where $S_y$ is his private key and, then, sends Y to the server.

The server and the client calculate the point $Z = S_x.Y = S_y.X = S_x.S_y.P$. Z has the form of $Z(z_x,z_y)$ where $z_x$ is the abscissa and $z_y$ is the ordinate. The shared key Ks can be $z_x$ or $z_y$ with a 128 bits length.

After this step, the PKG will generate the client private key $K_{priv}$ using his security parameters and the client YouTube address (as a unique identity). The public PKG security parameters are: E(p,A,B), p and $P_{pub} = s.P$ where s is the PKG private key. There is also a hash function called MapToPoint MTP which convert a simple string into a point in E(p,A,B). Then, the client's public key is MTP (client YouTube address) and $K_{priv}$ = s.MTP(client YouTube address).

The server sends the $K_{priv}$ and the Initialisation Vector (IV) encrypted with the shared key Ks.

Now, each client has a secure IV to start his scenario of accessing the server as shown in Fig. 9 or Fig. 10.

## IV. ATTACKS ANALYSIS

As our hashing technique uses keyed-hash functions, the majority of attacks can be thwarted. This part gives an overview on some high level attacks that can affect on the hashing or the link of chain. Those attacks may help in broking the hash link and causing some missing of video synchronization.

- Replay attack: (the attacks produced by delaying or deleting some video packets and resend them or anther to the destination along the same path). The *time-stamp* property can eliminate this attack. The Window transmission timestamp can resolve this problem by checking in the receiving side.
- Padding attack: (the attacks generated from adding some bytes to the original data and recalculate and resend the hash of new padded data). This attack can be easily eliminated by pre-pending the *Window Size or Length* because it is impossible to pre-pending the whole video length in case of online case but it is possible in offline streaming.
- Packet loss problems: The UDP transmissions are unreliable and cause some packet loss and others come within different order. The *indexing mechanism* of the *Window algorithm* can overcome on this problem.
- Collision attacks: this attack relevant to the hash algorithm used. MD5 and SHA-1 suffer from this attack which includes two aspects (Preimage and Birthday attacks). But, our hash construction can overcome these attacks as the following:
  - For Preimage attack: the RC calculation based concatenated 2, 3 or 4 Digest values gives impossibility to this attack. The concatenation gives some strongest to the final hash value. More over, the secure IV used will add some complexity to cryptanalysis attackers' procedures.
  - For birthday attack: the Multilayer construction increases the complexity of finding two messages having same Digest value.

Moreover, the hashing structure plays an important role in the degree of security. For using keyed-hash over unkeyed-hash have the following pros:

Keyed hash mechanism proposed in [21, 22] which called HMAC is a good example for Keyed-Hashing for Message Authentication Codes based on MD5 or SHA-1. It depends on secure shared key used with any standard cryptographic hash function between two parities to add some security measure for the message integrity and source-destination authentications. The degree of security could be increased if we used secure initialization vectors for hashing the windows of video stream. This IV can be created and defined by the same manner explained in [18] based on PKG private key generation system with the elliptic curve secured manner. Therefore, the value added to cryptographic hash functions by the keying system used can overcome many weakness and some attacks related to normal hashing or what is called unkeyed hash.

## V. RESULTS

In general, we built our analytical and simulation results based on the assumed parameters and values in Table I. Moreover, these assumptions were assumed based on some standards like packet size equal MTU and the delay times for video streaming within 1 to 2 sec maximum. But, it is important again to re-mention the difference in structure between *Packet*, *bock*, *Block* and *Window* as the following:

- **packet:** is the standard packet size 1500 Bytes
- **block:** is the standard size of block used by hash algorithm which is 512 bits for MD5 or SHA-1
- **Block:** is the number of packets to be processed together
- **Window:** is the total buffer which consists of number of Blocks depending on some parameters like: video rate, processing delay and RTT value

The Round Trip Time (RTT) is delay time consumed by the client to join the streaming server. It is important for our proposed algorithm to have total Delay time based on RC calculations and buffering or de-jittering less than the RTT or the client will prefer to initiate the session by requesting initialization vector. In this case the total Delay may be greater than RTT. The following equation expresses the total Delay related to buffering based static Window calculations:

$$D_{buffer} = D_{dejitter} + D_{calcul}$$

We have simulated the Redundancy Code Synchronization Recovery State (RC-SRS) algorithm and analyse some preliminary results. The obtained results based on some videos assumptions. Assume that, we have video file that needs buffer size (Bs) equal 2Mbit, then (Bs=2Mbits) and transmitting rate (R=1Mbit/Sec) then the total delay = (Bs/R) = 2 sec. So, if we have two delay times as the following:

$D_{calcul}$ : processing time for calculate hash (sender) and verification (receiver).

$D_{wait}$ : total delay time before starting using the RC to recalculate the hash link of any Block inside the Window according to the Block order in the Window (Best or Worst case as explained in Section III-D).

Then $D_{calcul}$+$D_{wait}$ must be less than (Bs/R) which 2 sec. So, our threshold condition will be:

$$(D_{calcul} + D_{wait}) < 2 \text{ sec}$$
$$D_{wait}=(\text{No packets x PS Packet Size})/ R \text{ (bits/sec)}$$

We will put the two sec in this case as Max threshold allowed delay time and change the number of packets to find the max number of Block size or buffer under the above conditions.

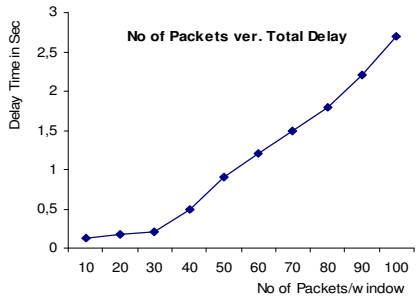

Fig. 12: The optimal number of packets under maximum allowable delay time (1 to 2 sec)

Fig. 12 shows the relation between the numbers of packets/Window or Block versus our assuming delay time from 1 to 2 sec. The curve gives 85 packets as optimum number for Window or Block size.

Fig. 12 compares the total overhead (the added bytes to stream as a redundancy code) by using MD5 or SHA-1 hash algorithms. As shown, if we assume the number of packets per block equal 10, so the full redundancy means sending the RC 10-times (means with each packet). But, this will lead to very high overhead.

If we take an example; for MPEG-TS the output chunks are equal 188 Bytes/chunk. The RTP packet size as we captured from the packet analyzer during the simulation was 1370 Bytes which equal 7(chunks) x188 Bytes + 54 (total headers rest). So, on the base of 1500 Bytes standard packets we still have 1500-(1370+2 bytes for index to the place of RC) = 128 Bytes. Those 128 Bytes give us the probability of sending the RC 8 times in case of MD5 and 6 times in case of SHA-1 as shown in Fig. 13. Those results were obtained under our assumptions of packet size 1500 Bytes and Block size 85 packets.

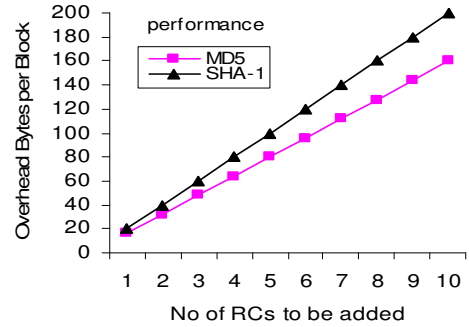

Fig. 13: The overhead bytes in terms of number of redundancy trials

In Fig. 14, a comparison between different methods has been made in terms of processing time for each Block of video against different video rates. This Block is almost is almost 2 sec in case of SHHC technique and 85 packets (for each packet size 1500 Bytes) in case of TSP, MLHC, TSP and RC techniques. The results indicate minimum accepted calculation time for our proposal based RC which the average time about 200 msec for each Block.

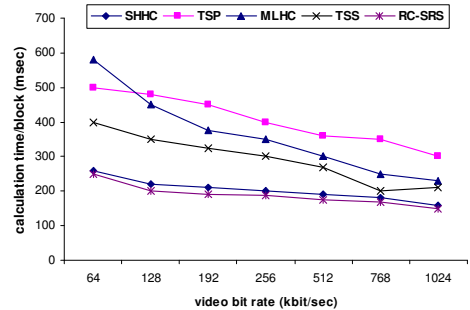

Fig. 14: The computation time comparison between the different methods against different videos bit rates.

For the error rate and its impact on hash recovering or reliability of our technique we have two parameters control this process:

*Packet Error Rate (PER)*: is the probability of an error occurred in any packet of the *Block*. So,

$$PER=1/\text{Block size}$$

If the *Window* has (M) *Blocks* then the total will be:

$$PER_T=(1/\text{Block size})/M$$

*Hash Error Rate (HER)*: is the probability of an error occurred exactly in the packet which carried the RC:

$$HER= PER_T/\text{Block size}$$

This calculation is valid for the case of each *Window* contains one *Block*. But, if we have different Redundancy Factors (RF) like 2/3, 3/4 and 4/5 then, we must multiply the HER by the RF as:

$$HER= (PER_T/\text{Block size}).RF$$

TABLE II
OVERHEAD COMPARISON

| Resynchronization Technique | The Method Used | MD5 16-Bytes /window | SHA-1 20-Bytes/window | Overhead processing Delay |
|---|---|---|---|---|
| SHHC | -Hashes Concatenation (3 hashes) <br> -Hashes XOR | 48-Bytes/window <br> 16-Bytes/window | 60-Bytes/window <br> 20-Bytes/window | 3.X |
| TSP | One packet/window | N-Bytes | N-Bytes | X |
| MLHC | 2 Layers hash | 16-Bytes | 20-Bytes | ≈2.X |
| TSS | 4 Bytes/packet for timestamp | 20-Bytes | 24-Bytes | X |
| RC-SRS | -Redundancy 2/3 <br> -Redundancy 3/4 <br> -Redundancy 4/5 | 16-Bytes <br> 16-Bytes <br> 16-Bytes | 20-Bytes <br> 20-Bytes <br> 20-Bytes | 2.X <br> 3.X <br> 4.X |

In this comparison N is the packet size and X is the processing time for each *Block/Window* buffers and X<<RTT.

In Fig. 15, the performance of the hybrid technique based RC in terms of probability of recovery against different packet loss error rates has been illustrated. The simulation has been done using SHA-1 hashing technique. The three redundancy factors used are 2/3, 3/4 and 4/5 (which means combines 2 hashes, 3 hashes or 4 hashes values per static *Window*) give high probability of chain recovery till error rate 0.2 % for all RF. This gives us good indication for the robustness of our proposal and its high degree of recovery stat of chain link.

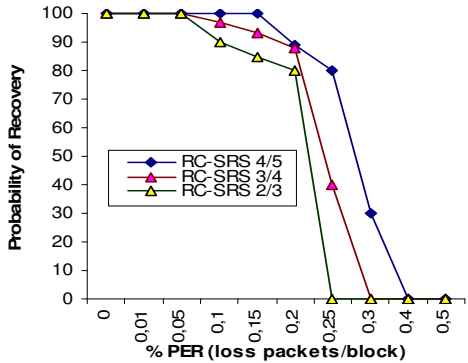

Fig. 15: The percentage of recovery state against the packet error rate PER for video rate 1024 kbps with different redundancy factors 2/3, 3/4 and 4/5.

Table II summarizes the behaviour of each algorithm by comparing between them in terms of overhead and processing delay per each Window or Block. The overhead for the hybrid technique based RC is almost the same overhead of the others or less than them. Also, the max delay time is 4.X is less than RTT value for the clients.

### A. Advantages of Small Window or Block Size

As the *Window* size is varied based on the video rate, the achieving degree of reliability has some how relation with the increasing of *Window* size. It is best to keep the *Window* as small as possible so as to enhance the reliability of transmission. This is very important in transmission over UDP because its nature is unreliable by default. But, for TCP as it is reliable transmission so, the impact of *Window* increasing will not have great effects on the reliability.

Although many previous works simulated the large number of packets per Block, our work has good benefits from adopting small Window buffers like:

1. more reliable with unreliable transmission environment like UDP transport systems
2. fast calculation and verification time
3. small overheads
4. the Block with 85 packets seems a small size, but this assumption has a good features on PER or losses till complete Window
5. the HER will be controllable under these assumptions

### B. Adaptive Window Size

This work adopts static sliding Window which means that; each Window consists of fixed number of Blocks. The Window size is negotiated between the server and client according to the video rate and the client capabilities during the session establishment. The client will process the Window for specific number of Blocks then empty it. So, there is no relation between the current Window and the next one else the hash value of the last Block of this Window. So, the receiver only cashes small information from the previous window which is its digest value to use it as initialization vector for the next one.

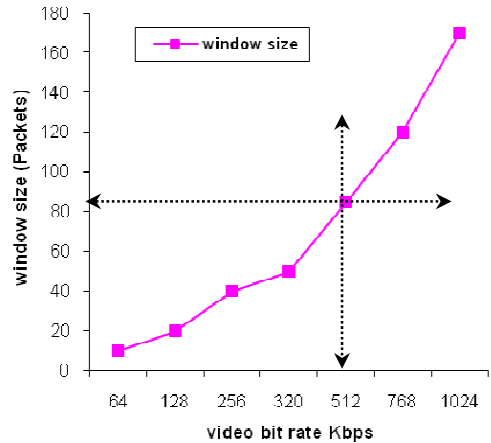

Fig. 16: window size for different video rates against video rates (64K, 128K, 256K, 512K, 768K and 1M)

As we agree before from Fig. 12 that we build our comparison on *Window* or *Block* size 85 packets for video bit rate 1024 Kbps and for a delay time 1 sec, this buffer will be duplicated if we assume 2 sec delay as shown in Fig. 16. In this figure, there is a comparison between different video bit rates against the elevation effects of *Window* size. As the rate increased the window size must be increased and vice verse. But, decreasing the video rate below 512 Kbps will lead us to *Window* size less than 85 packets which by default has negative impact on the delay time augmentation.

## VI. CONCLUSION

This work focused on the resynchronization needs for hash chain mechanisms in video security by using redundancy codes (RCs). Therefore, we categorized the methods that can be used in hash chain link recovery into four (SHHC, TSP, MLHC and TSP). Then, we proposed a hybrid technique called RC-SRS that inherits from the pros of all previous techniques.

In terms of complexity, a comparison has been made between the different ways for achieving the resynchronization of hash chain. This is followed by an evaluation of our proposed method RC-SRS for resynchronisation based on redundancy codes and a study on attack mitigations. Our results indicted that; the RCs will not cause additional computation time for the sender and receivers and the overhead added is accepted in terms of packet size. Moreover, the delay time consumed by the receiver to deduce the hash link based on received RC is less than standard RTT.

As, this work adopted static sliding *Window* technique for calculating the RCs, our prospection is to simulate the dynamic case. This scenario will be built using dynamic sliding *Window*. We expect that this scenario will add more robustness besides increasing the degree of reliability and the degree of recovery.

## REFERENCES

[1] H.Aslan; 'A hybrid scheme for multicast authentication over lossy networks'; Computers & Security (2004) 23, pp.705-713.
[2] A.Perrig, R.Szewczyk, V.Wen, D.Culler, J. D. Tygar; 'SPINS: Security Protocols for Sensor Networks', ACM Mobile Computing and Networking, Rome, Italy, 2001.
[3] I.Syamsuddin; T.Dillon; E.Chang; H.Song; 'A Survey of RFID Authentication Protocols Based on Hash-Chain Method'; Third International Conference on Convergence and Hybrid Information Technology ICCIT 08; Vol. 2, 11-13 Nov. 2008, pp.559 – 564.
[4] H.Guo , Y.Li , S.Jajodia; 'Chaining watermarks for detecting malicious modifications to streaming data'; Information Sciences 177 (2007),pp.281–298.
[5] M-S.Hwang and P.Sung ; 'A Study of Micro-payment Based on One-Way Hash Chain'; International Journal of Network Security, Vol.2, No.2, Mar. 2006, pp.81–90.
[6] L.Lamport; 'Password authentication with insecure communication'; Communications of the ACM, 24(11): November 1981, pp770-772.
[7] Y.LIU, L.HU, and H.LIU;' Using an efficient hash chain and delaying function to improve an e-lottery scheme '; International Journal of Computer Mathematics; Vol. 84, No. 7, July 2007, pp.967–970.
[8] Y.Challal, A.Bouabdallah and Y.Hinard; 'RLH: receiver driven layered hash-chaining for multicast data origin authentication'; Computer Communications 28 (2005), pp.726–740.
[9] R.Gennaro and P.Rohatgi; 'How to sign digital streams', In Proceedings of the Advances in Cryptology CRYPTO'97, 1997, pp.180-197.
[10] Z.ZHISHOU, J.APOSTOLOPOULOS, Q.SUN, S.WEE and W.WONG; 'Stream authentication based on generalized butterfly graph'; In Proceedings of the IEEE International Conference on Image Processing (ICIP'07), Vol. 6. 2007, pp.121–124.
[11] A. Perrig, R. Canetti, J. Tygar and D. Song, "Efficient authentication and signing of multicast streams over lossy channels," in Proc. of IEEE Symposium on Security and Privacy, 2000, pp. 56-73.
[12] P. Golle and N. Modadugu, "Authentication streamed data in the presence of random packet loss," ISOC Network and Distributed System Security Symposium, 2001, pp.13-22.
[13] E.Abd-Elrahman; H.Afifi;' Optimization of File Allocation for Video Sharing Servers', NTMS 3rd IEEE Conference, Dec. 2009, pp.1 – 5.
[14] Y. Kikuchi, T. Nomura, S. Fukunaga, Y. Matsui, H. Kimata ; 'RTP Payload Format for MPEG-4 Audio/Visual Streams', RFC 3016, Nov. 2000.
[15] J. Staddon, S. Miner, M. Franklin, D. Balfanz, M. Malkin and D. Dean, ''Self-healing Key Distribution with Revocation'', proceedings of IEEE Symposium on Security and Privacy, pp. 224-240, 2002.
[16] National Institute of Standards and Technology (NIST), 2002, FIPS 180–2: 'Secure hash standard' (Washington, DC: NIST, US Department of Commerce).
[17] E.Abd-Elrahman, M.Boutabia and H.Afifi, 'Video Streaming Security: Reliable Hash Chain Mechanism Using Redundancy Codes', The ACM 8th International Conference on Advances in Mobile Computing and Multimedia (MOMM 2010) Paris, France, 8-10 Nov 2010, pp.69-76.
[18] E. Abd-Elrahman, M. Abid and H.Afifi, '' Video Streaming Security: Window-Based Hash Chain Signature Combines with Redundancy Code'', The IEEE International Symposium on Multimedia (ISM2010) Taiwan, Dec 2010, pp.33-40.
[19] L.Ronald, R.Rivest. The MD5 message-digest algorithm. Internet Request for Comment RFC 1321, Internet Engineering Task Force, April 1992.
[20] National Institute of Standards and Technology (NIST). Secure Hash Standard (SHA), Federal Information Processing Standards (FIPS) Publication 180-1, May 1993.
[21] H. Krawczyk, M. Bellare, R. Canetti; ''HMAC: Keyed-Hashing for Message Authentication'', Request for Comments: 2104, Feb 1997.
[22] American Bankers Association, Keyed Hash Message Authentication Code, ANSI X9.71, Washington, D.C., 2000.
[23] M.Hefeeda and K.Mokhtarian; ''Authentication Schemes for Multimedia Streams: Quantitative Analysis and Comparison'', ACM Transactions on Multimedia Computing, Communications and Applications, Vol. 6, No. 1, Article 6, February 2010.
[24] M.-J. Montpetit, G. Fairhurst, H. Clausen, B. Collini-Nocker, H. Linder ;'' A Framework for Transmission of IP Datagrams over MPEG-2 Networks'' , RFC 4259, Nov. 2005.
[25] X. Wang and H. Yu, "How to Break MD5 and Other Hash Functions", Proceedings of EuroCrypt 2005, Lecture Notes in Computer Science, Vol. 3494, 2005.
[26] NIST Special Publication (SP) 800-107, Recommendation for Applications Using Approved Hash Algorithms, July 2008.
[27] Q.Wang, H.Khurana, Y.Huang, K.Nahrstedt, ''Time Valid One-Time Signature for Time-Critical Multicast Data Authentication'', IEEE INFOCOM 2009, pp.1233-1241.
[28] E. Rescorla, ''Diffie-Hellman Key Agreement Method'', RFC 2631, June 1999.
[29] S. Blake-Wilson, N. Bolyard, V. Gupta, C. Hawk, B. Moeller; 'Elliptic Curve Cryptography (ECC) Cipher Suites for Transport Layer Security (TLS)', RFC 4492, May 2006.
[30] A.Shamir, ''Identity-Based Cryptosystems and Signature Schemes'', 1984.
[31] Zhishou Zhang; Qibin Sun; Wai-Choong Wong; Apostolopoulos, J.; Wee, S.; , "Rate-Distortion-Authentication Optimized Streaming of Authenticated Video," *Circuits and Systems for Video Technology, IEEE Transactions on* , vol.17, no.5, May 2007, pp.544-557.
[32] Qibin Sun; Zhi Li; Yong Lian; Chang Wen Chen; , "Joint Source-Channel-Authentication Resource Allocation for Multimedia overWireless Networks," *Circuits and Systems, 2007. ISCAS 2007. IEEE International Symposium on*, 27-30 May 2007, pp.3471-3474.